\documentclass[12pt,preprint]{aastex}

\usepackage{epsfig}
\usepackage{amsmath}
\usepackage{subfigure}
\usepackage{natbib}
\usepackage{multirow}
\usepackage{txfonts}
\begin{document}
\title{Discovery of Time Variation of the Intensity of Molecular Lines in IRC+10216 in The Submillimeter
and Far Infrared Domains}

\shorttitle{Infrared pumping of molecular lines}
\shortauthors{Cernicharo et al.}

\author
{
J. Cernicharo\altaffilmark{1}, 
D. Teyssier\altaffilmark{2},
G. Quintana-Lacaci\altaffilmark{1}
F. Daniel\altaffilmark{3},
M. Ag\'undez\altaffilmark{1},
L. Velilla Prieto\altaffilmark{1},
L. Decin\altaffilmark{4},
M. Gu\'elin\altaffilmark{5},
P. Encrenaz\altaffilmark{6},
P. Garc\'\i{}a-Lario\altaffilmark{2},
E. de Beck\altaffilmark{7},
M.J. Barlow\altaffilmark{8},
M.A.T. Groenewegen\altaffilmark{9},
D. Neufeld\altaffilmark{10}
J. Pearson\altaffilmark{11}
}

\altaffiltext{1}{Group of Molecular Astrophysics. ICMM. CSIC. C/Sor Juana In\'es de La Cruz N3. E-28049, 
Madrid. Spain}
\altaffiltext{2}{ESA. ESAC. P.O. Box 78, Villanueva de la Ca\~{n}ada. E-28691 Madrid. Spain}
\altaffiltext{3}{Univ. Grenoble Alpes, IPAG, F-38000 Grenoble, France 
CNRS, IPAG, F-38000 Grenoble, France}
\altaffiltext{4}{Instituut voor Sterrenkunde, Katholieke Universiteit Leuven, Celestijnenlaan 200D, B-3001 Leuven, Belgium}
\altaffiltext{5}{Institut de Radioastronomie Millim\'etrique, 300 rue de la Piscine, F-38406, St-Martin d'H\`eres, France}
\altaffiltext{6}{LERMA, Observatoire de Paris, 61 Av. de l'Observatoire, F-75014 Paris, France}
\altaffiltext{7}{Department of Earth and Space Sciences, Chalmers University of Technology, Onsala Space Observatory, SE 439 92 Onsala, Sweden}
\altaffiltext{8}{Department of Physics and Astronomy, University College London, Gower Street, London WC1E 6BT, UK}
\altaffiltext{9}{Koninklijke Sterrenwacht van Belgi\"e, Ringlaan 3, B-1180, Brussel, Belgium} 
\altaffiltext{10}{Department of Physics and Astronomy, Johns Hopkins University, 3400 North Charles Street, Baltimore, MD 21218, USA}
\altaffiltext{11}{Jet Propulsion Laboratory, California Institute of Technology, Pasadena, CA, 91109, USA}

\begin{abstract}
We report on the discovery of strong intensity variations in the high rotational lines
of abundant molecular species towards the archetypical circumstellar envelope of IRC+10216. The observations have
been carried out with the HIFI instrument on board \textit{Herschel}\thanks{\textit{Herschel} is an ESA space observatory with 
 science instruments provided by European-led Principal Investigator consortia and with 
 important participation from NASA}
 and with the 
IRAM\thanks{This work was based on observations carried out with the 
IRAM 30-meter telescope. IRAM is supported by INSU/CNRS (France), 
MPG (Germany) and IGN (Spain)} 30-m
telescope. They cover several observing periods spreading over 3 years. The
line intensity variations for molecules produced in the external layers of the envelope most 
probably result from time variations in the infrared pumping rates.
We analyze the main implications this discovery has on the interpretation of molecular
line emission in the envelopes of Mira-type stars. Radiative transfer calculations have
to take into account both the time variability of infrared pumping
and the possible variation of the dust and gas temperatures with stellar phase
in order to reproduce the observation of molecular lines at different epochs. 
The effect of gas temperature variations with stellar phase
could be particularly important for lines produced in the innermost regions of the envelope.
Each layer of the circumstellar envelope sees the stellar light radiation with a different
lag time (phase). Our results show that this effect must be included in the models. 
The sub-mm and FIR lines of AGB stars cannot anymore be considered as safe intensity calibrators.
\end{abstract}

\vspace{0.5cm}
To appear in the Astrophysical Journal Letters
\vspace{0.5cm}

\date{Received Oct 7 2014; accepted Oct 21 2014}

\section{Introduction}
Half of the known interstellar molecular species are detected in IRC+10216, the envelope of the AGB star 
CW Leo and one of the brightest infrared sources in the sky. CW Leo, at a estimated distance of
$\simeq$130\,pc from the Sun, is a Mira variable star with a period
of 630-670 days and an amplitude of $\simeq$1 mag in the $K$ band 
\citep[][and references therein]{Menten2012}. 

Due to its proximity, its large mass loss rate and wealth of molecules, IRC+10216 attracted much studies 
and is considered as the archetype of high-mass loss AGB stars. Yet, the formation of the dust and of 
molecules, some of which are fairly complex, and their dependance on the stellar state of evolution, 
from AGB to pre-PN, are key issues not yet fully understood.
The \textit{Herschel} satellite \citep{Pilbratt2010} recently allowed much progress in this domain. The 
stellar atmosphere and 
the dust condensation zone are best probed by ro-vibrational lines 
and by high-$J$ pure rotational lines pertaining to the ground and vibrationally excited states. Those lines 
lie in the infrared and far infrared (FIR) for the most abundant species, such as CO, HCN and 
SiS \citep{Fonfria2008,Cernicharo1996,Cernicharo2010a,Decin2010}. 
In contrast, the outer cold layers of the envelope 
are best probed by low-$J$ rotational lines at mm and sub-mm wavelengths. 

The determination of the physical conditions througout the envelope can be performed
by combining a large set of FIR, mm, and sub-mm data.
However, and despite the fact that IR pumping is recognized to play a role in the excitation of 
molecular lines in evolved stars (\citet{Agundez2006} for H$_2$O; \citet{Deguchi1984} for HC$_5$N); 
\citet{Daniel2012} for HNC; \citet{DeBeck2012} for CCH), most studies carried out so far assumed 
steady state, i.e., that the known periodic variations of the stellar IR flux do not modulate 
molecular line emission. This was partly justified by lack of evidence of line intensity variations, 
outside a few cases of strong maser emission in O-rich AGBs for which maser line and IR continuum 
intensities are clearly correlated \citep{Pardo2004,Nakashima2007}.
\citet{Cernicharo2000}, for example, unsuccessfully searched for line intensity variations 
in their 2-mm spectral survey of IR+10216, which combines observations scattered over 12 years and 
mostly detects low-$J$ transitions from fairly abundant species and their isotopologues.

\begin{figure*}
\begin{center}
\includegraphics[angle=0,scale=.80]{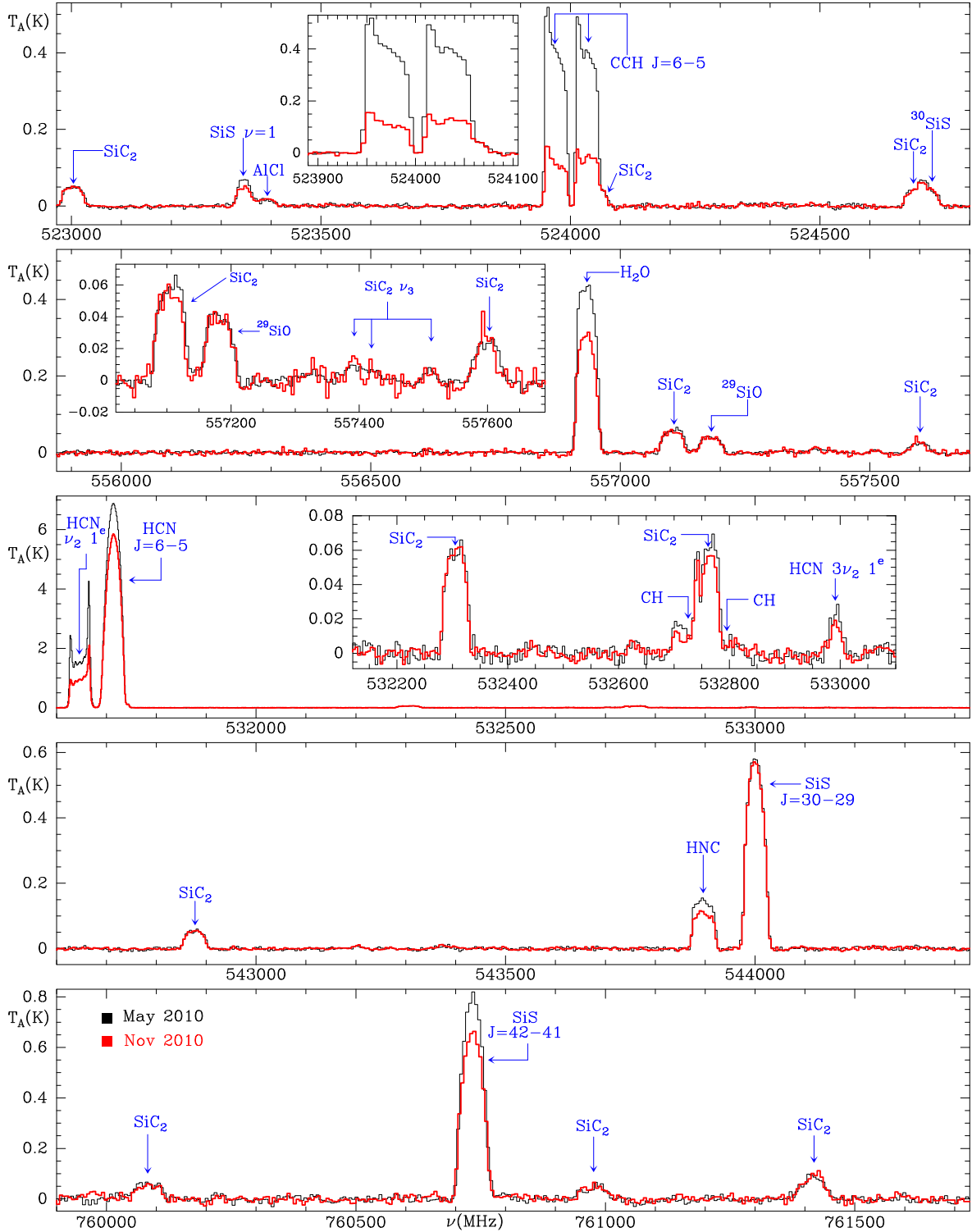}
\caption{Selected spectra observed with \textit{Herschel} in May 2010 (black thin
spectra; from \citet{Cernicharo2010b}) and November 2010 (red thick spectra).
The spectrum at 543.5 GHz was taken in May 2011.
Windows within the panels show zooms to selected lines. Intensity scale is antenna
temperature in K and abcissa is frequency in MHz.} 
\label{Fig1}
\end{center}
\end{figure*}

We present the results of a time monitoring of molecular line thermal emission in 
IRC\,+10216 taken between 480 and 1907 GHz with the \textit{Herschel/HIFI} instrument over 3 years,  
a time interval longer than the light period of IRC+10216 and with the IRAM 30-m telescope 
in a selected sample of frequencies between 85 and 350 GHz. 

We focus in this Letter on the study of CCH ($N$=1-0, 3-2, 4-3, 6-5, 7-6, 8-7)
and HNC ($J$=1-0, 3-2, 6-5, 7-6, 8-7). We present the first evidence for strong time 
variation of the intensity of those lines, as well as of high excitation lines of several 
other molecular species. 

\begin{figure}
\begin{center}
\includegraphics[angle=0,scale=.62]{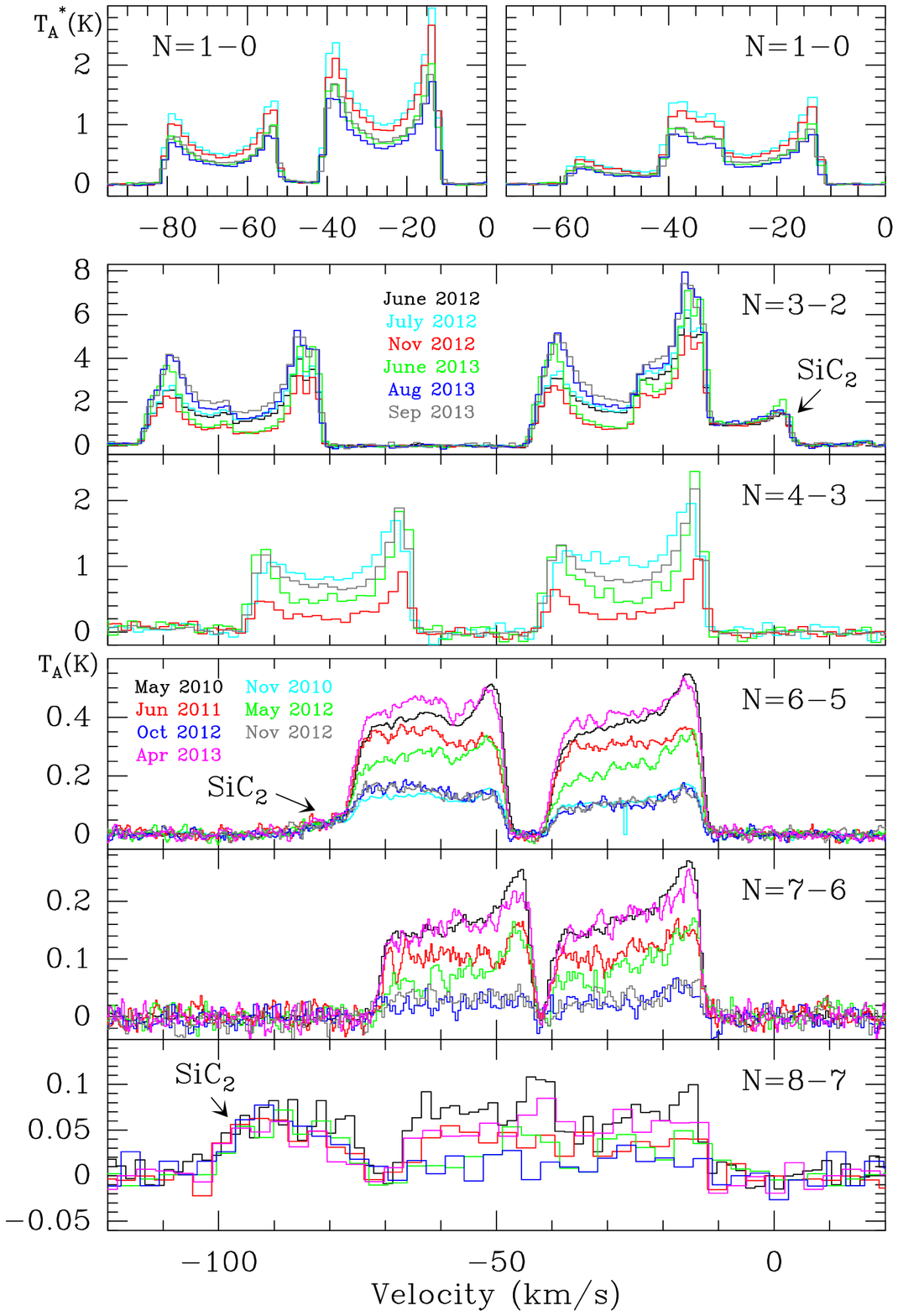}
\caption{CCH lines observed with the IRAM 30-m telescope ($N=1-0, 3-2, 4-3$) 
and with HIFI ($N=6-5, 7-6, 8-7$). The different colors correspond to the observing
dates with both instruments. The abscissa is velocity in km\,s$^{-1}$ with respect the
frequency of the strongest hyperfine component. Intensity scale is antenna
temperature in K.}
\label{Fig2}
\end{center}
\end{figure}

\section{Observations and Results}
A line survey of IRC+10216 was carried out on May 11-15 2010 with HIFI \citep{deGraauw2010}
using all HIFI bands between 480 and 1907\,GHz. 
These data have been previously presented by \citet{Cernicharo2010b}. 
%The single side band spectrum of the whole line survey was obtained through a standard manual 
%procedure as decribed by \citet{Cernicharo2010b}.
%The calibration and the double side band gain ratio was checked against the strongest lines, and 
%found to be consistent 
%within 2-4\% between upper and lower frequency settings, and between the H and V receivers of HIFI. 

A second set of data
was taken during November 2010 to search for light hydrides in this source
\citep[HCl, HF, PH$_3$,][]{Cernicharo2010c,Agundez2011,Agundez2014}. A total of 66 frequency
settings with 4+4 GHz bandwidth (USB and LSB) were acquired covering a large number of lines
observed in both periods. Some frequency settings for that proposal were taken in May 2011.
Both data sets have an rms noise of 7-12\,mK in 3 km\,s$^{-1}$-wide spectral channels.

The comparison of the lines taken at six month interval revealed surprisingly large 
intensity variations, as illustrated on Figure 1. These variations, which, reach a factor of 
3 in some cases, are well above any possible instrumental effect: pointing errors, 
gain calibration, sideband imbalance, baseline ripples, spurious ghost lines, etc.
We found, in particular, that:

\begin{figure}
\begin{center}
\includegraphics[angle=0,scale=.62]{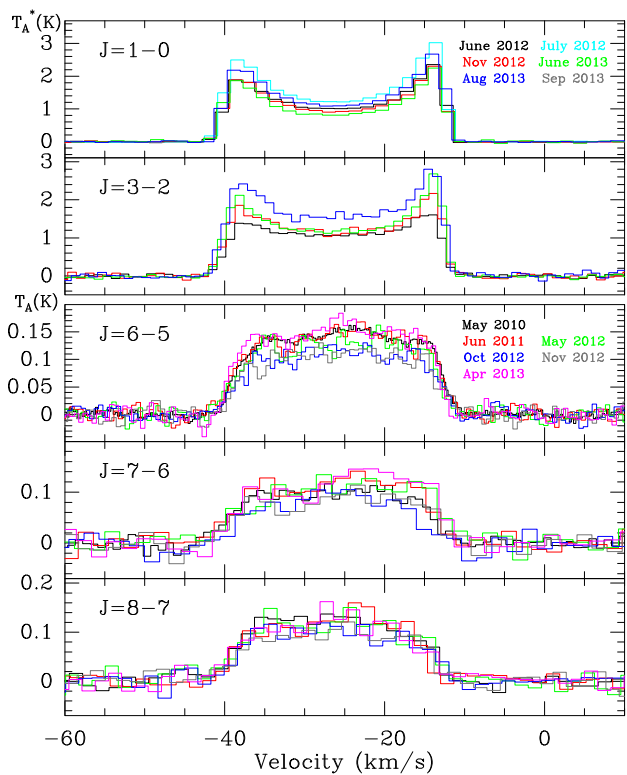}
\caption{HNC lines observed with the IRAM 30-m telescope ($J=1-0, 3-2$) 
and with HIFI ($J=6-5, 7-6, 8-7$). The different colors correspond to the observing
dates with both instruments. Intensity scale is antenna
temperature in K and the abscissa is velocity in km\,s$^{-1}$}
\label{Fig3}
\end{center}
\end{figure}

\vspace{0.2cm}
\noindent
1) More than 95\% of the lines of SiC$_2$ observed in both periods have exactly the same intensity. 
\vspace{0.2cm}

\noindent
2) Spectra with lines exhibiting an intensity variation also contain many other lines, 
detected with very good signal to noise ratio ($>$10), that do not show any intensity 
fluctuation. 
\vspace{0.2cm}

\noindent
3) CS, SiO and SiS show a differential effect. Low$-J$ transitions  have  low   
line  intensity changes. However, this variation is larger with large $J$ reaching a value $\simeq$25\%  
for the three species (see last two panels of Figure 1 for SiS). 
\vspace{0.2cm}

\noindent
4) The above variations are not directly related to the source extent, as they are observed both for 
SiS, which is 
spatially constrained near the star, and to CCH, which is only observed in the outer envelope. 
The CCH $N=6-5$ through $8-7$ lines show a factor of 3 of change in intensity, whereas 
two SiC$_2$ lines, parly blended with the high CCH fine structure component remain constant.   
\vspace{0.2cm}

\noindent
5) Water vapor shows a variation of 50\%. Following the models of \citet{Agundez2006},  
which include infrared pumping of H$_2$O, this result could had been expected.
\vspace{0.2cm}

\noindent
6) HNC shows a systematic change of 20-50\% in the integrated
intensity of all its lines observed with HIFI.
\vspace{0.2cm}

\noindent
7) The CO lines observed at both epochs have low-$Js$ and 
show no significant intensity variation. 
However, the $J=17-16$ line of $^{13}$CO, the only$^{13}$CO line observed 
in both periods, appears to vary in intensity by 20\%.

\vspace{0.2cm}

\noindent
8) 
HCN lines below $J=14-13$ have strong changes, ranging from 20-50\%.  
HCN  is  a  peculiar  case,  with  several  regimes  
in  the  excitation  of  its  ro-vibrational  levels. 
This makes the excitation diagram quite complex: for example, in  the  inner  
shells,  the $\nu_2$  bending  state  
has a population reaching as much as 30\% of that of the ground state 
\citep{Cernicharo1996,Cernicharo2010a}. 
The modelling  of  the  observed  
HCN emission  requires the inclusion of vibrational levels  up  to  12000  K 
\citep{Cernicharo1996,Cernicharo2010a,Cernicharo2013}. 
\vspace{0.2cm}

\begin{figure}
\begin{center}
\includegraphics[angle=0,scale=.62]{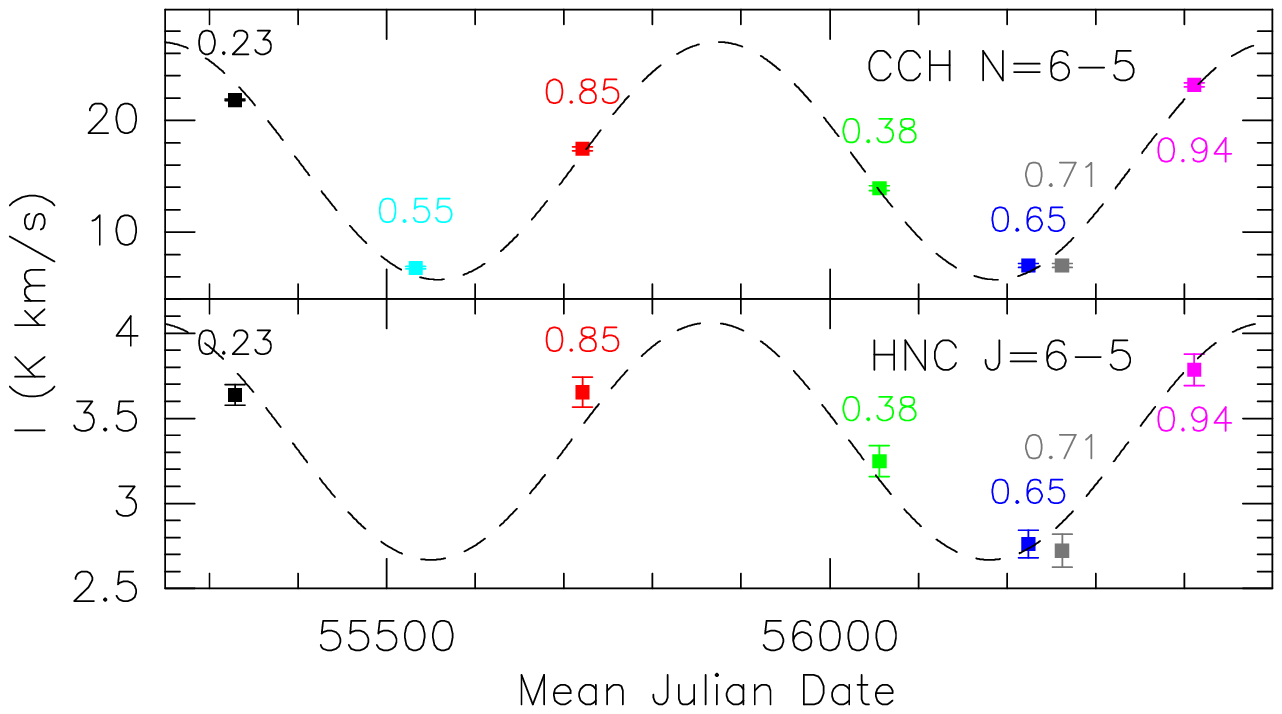}
\caption{CCH (top panel) and HNC (bottom panel) $N/J=6-5$ 
integrated line intensity for 
each epoch using the same colour code as in Figures 2\&3. 
Error bars correspond to the 1\,$\sigma$  noise rms of the spectra.
The dashed sine curve (period of 630 days)
corresponds to the stellar light curve reported by \citet{Menten2012}.
The stellar phase is indicated for each observing date.
}
\label{Fig4}
\end{center}
\end{figure}

Prompted by these unexpected results, we succeeded in  extending by 3 more years our 
IRC+10216 study with \textit{Herschel}. In total, 7 observing runs extending from May 
2010 to May 2013 were made with HIFI, using the same observing procedure in all runs. 
In addition, complementary lower frequency observations were carried out with the IRAM 
30-m telescope between 2012 and 2013. Note that the observing periods were similar, but 
not identical on both instruments due 
to scheduling and weather constraints (we only kept data taken when the 
precipitable water atmospheric content was between 1 and 10 mm). 
The system temperatures werethen in the range 80--300 K and the 
pointing errors below 3$''$. 

In this work we will focus on the lines of CCH and HNC throughout 2010-2013.
The May 2010 data, taken during the first HIFI line survey, 
have already been analysed for CCH, HNC, CO and SiC$_2$ and published
\citep{Cernicharo2010b,Cernicharo2010c,Muller2012,DeBeck2012,Daniel2012}. 
Full results for all detected molecular species, including data from 
other $Herschel$ instruments will be published in a forthcoming paper. 

Figure 2 shows the CCH line profiles (composed of several hyperfine components -$hfs$-)
observed at different epochs and  Figure 3 shows those of HNC. For
both molecules, even the lowest $N/J$ rotational transitions exhibit time variations in intensity and shape. 
The amplitude of the CCH line intensity variations changes
with the rotational quantum number $N$, the higher-$N$, the larger the amplitude
(a factor of 3 for $N=6-5$ and about 7 for $N=7-6$). This is consistent with the expectation that collisional excitation is less 
efficient for the higher rotational levels of CCH, a molecule only present in the 
outer envelope, making infrared pumping relatively more efficient.
It is worth noting that the strongest $hfs$ component of CCH $N=3-2$ (Figure 2 second panel from top) 
is partially blended with a line of SiC$_2$ ($12_{0,12}-11_{0,11}$) which shows little or no 
intensity variation.  
The IRAM 30-m telescope observations show many other molecular lines with no 
significant intensity changes. Hence, the observed time fluctuations of the $N=1-0, 3-2, and 4-3$ lines
of CCH and $J=1-0, and 3-2$ of HNC are not produced by calibration or pointing errors.

Besides being definitively real, the CCH and HNC line intensity variations correlate well 
with the periodic stellar light variations. Figure 4 shows the observed line intensities, 
as derived from Figures 2\&3, atop the predicted stellar IR intensity (derived from 
\citet{Menten2012} for a period of 630 days) after applying shifts of respectively
51 days and 59 days for HNC and CCH. 
This shift 
%may account for the localisation of those species in the outer envelope 
%and 
is similar to that observed by \citet{Groenewegen2012} for the FIR dust continuum emission.
Infrared light variations, by modulation of radiative pumping of the molecular levels, and, 
possibly in the innermost layers of the CSE by modulating the dust grain and gas temperatures,
clearly appear as the cause of the observed molecular line intensity variations. 
These effect must definitively be included in radiative transfer ($RT$) models.

\begin{figure}
\begin{center}
\includegraphics[angle=0,scale=.58]{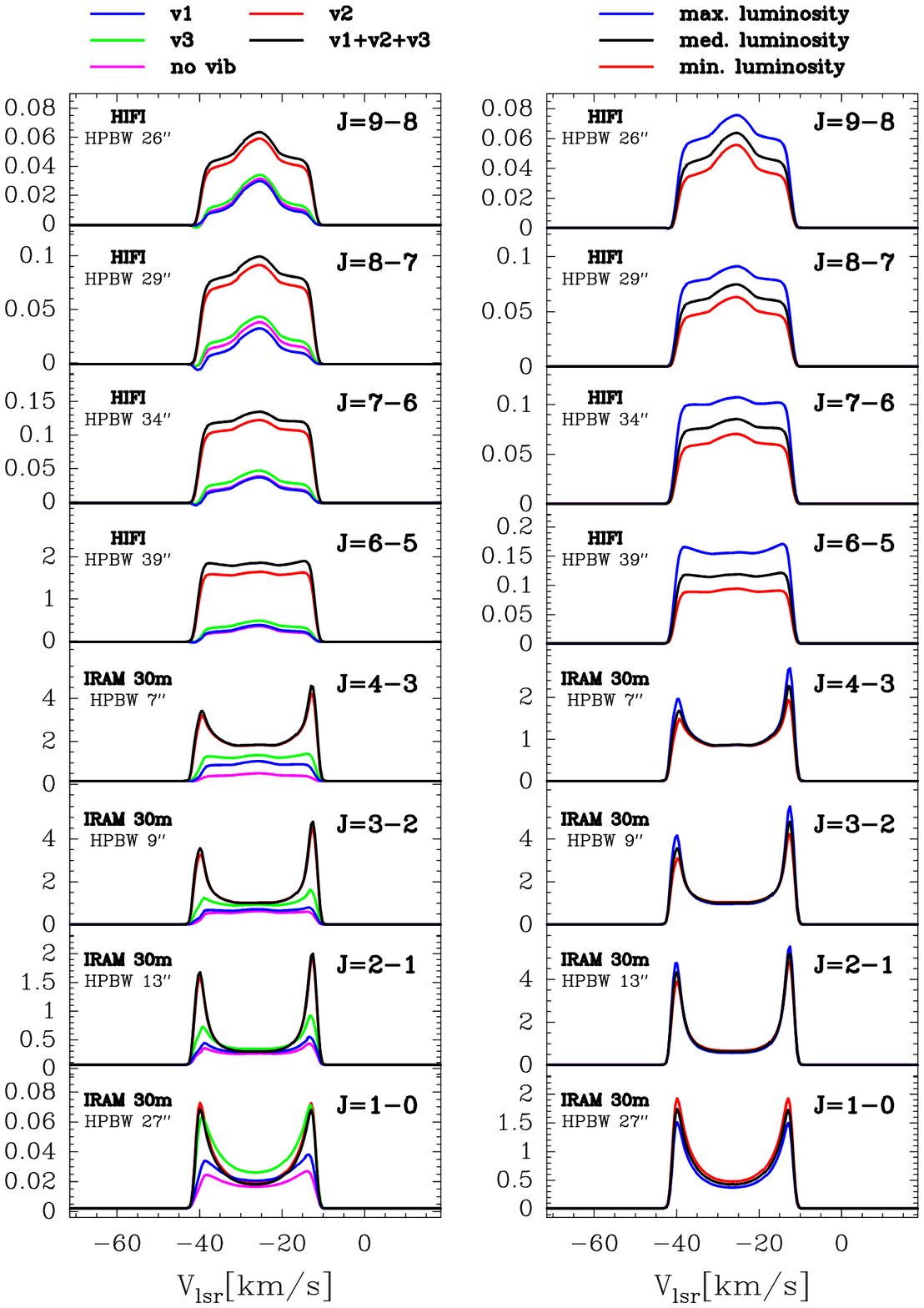}
\caption{{\it Left Panel:} Effect of the infrared pumping on the population of the rotational
levels of HNC for transitions from $J=1-0$ up to $J=10-9$. The different colors show the emerging
spectra when each one of the vibrational modes of HNC is included in the infrared
(as indicated above the top panel). The parameters adopted for the model are those of
\citet{Daniel2012} and \citet{Cernicharo2013}. The emerging profiles have been computed for different telescopes as
indicated in each panel. The stellar phase was assumed to be 0.3.
{\it Right Panel:} Emerging HNC profiles 
for three different stellar phases (as indicated above the top panel),
including infrared pumping through all vibrational modes.
%and the same physical parameters as for the left panels.
%The parameters for the model are those of \citet{Daniel2012} (see also \citet{DeBeck2012}).
Intensity scale is antenna temperature in K and abscissa is velocity in km\,s$^{-1}$ for both
panels.} 
\label{Fig5}
\end{center}
\end{figure}

\section{Discussion}
The results described above require a detailed and careful revision of the
interpretation and modelling performed up to now. Most of the
$RT$ models were aimed at deriving the 
population of the molecular energy levels of the abundant species, such as CO, HCN, SiO, SiS, 
CS, OH and H$_2$O. All these molecules are characterized by
thermal emission in their ground vibrational state. Nevertheless, some of them also
show strong maser emission and these maser transitions can be used only to derive 
some estimates of the physical parameters of the envelope.

The modelling of the emission of thermal lines relies 
on assuming values, or taking them from observations when possible, for the temperatures and sizes 
for the central star and the dust formation zone (the latter may extend up to 20 R$_*$ 
\citep{Fonfria2008}), as well as the
temperature and density profiles for the gas and the dust throughout the envelope. 
One generally assumes that the gas volume density varies as $n\simeq r^{-2}$) and that 
the geometry is spherically symmetric. One must also adopt an abundance profile for the 
key molecular species. The terminal expansion velocity in the envelope 
and the velocity field within the dust formation zone can then be derived from observations 
(see, e.g., \citet{Fonfria2008}, \citet{Cernicharo2010a}).

From the observation of a large number of molecular lines of a given molecule 
with widely different energy levels several of these parameters can be constrained in 
steady state. However, when the central star and dust formation zone show significant 
flux intensity variations, i.e. when the steady state does not hold anymore,
the problem becomes much more complex. 
The phase of the radiation field, seen at a given time by each point of the envelope, depends 
on the distance to the central radiation. Moreover, the temperature of the local gas, hence the 
collisional excitation rates, may also be modulated with different phase lags, since the gas 
heating and cooling are dominated in the inner regions by IR transitions 
that have large Einstein coefficients. The equilibrium times for the gas with the stellar and 
dust fluxes could there be fairly short.
Statistical equilibrium equations assume that 
dn$_i$/dt=0, where n$_i$ represents the population of level $i$. 
However, now the source function, which describes the radiation field reaching a point of the
envelope, depends on time in a complex way. Not only the radiation directly incoming 
from the star and its surrounding hot dust shell, comes at a given point P of the envelope 
with a time lag, as described above, but each envelope point radiatively connected to P sees 
this stellar radiation with a different phase, because it lies at different distance from the 
star and is distant from P. The magnitude of this effect on the $RT$ depends on the period of 
the star, the size of the envelope, the molecular abundance profile, the local density and the radiative transitions considered.
Hence, solving the $RT$ problem requires to couple statistical equilibrium equations
with the stellar light flux and the molecular emission arising from all points of the envelope, each point
seeing the stellar radiation with a different phase.

The role of infrared pumping of rotational lines has been discussed many times in the literature.
\citet{Agundez2006} have shown that including it for H$_2$O decreases the abundance 
of this molecule in IRC+10216 by a factor 10 as compared to a case in which only collisional 
excitation is considered. In triatomic
molecules such as CCH and HNC infrared pumping has been considered by \citet{DeBeck2012} and \citet{Daniel2012}.
CCH has a peculiarity when compared with HNC. 
The $\nu_2^1+2\nu_3$ and $\nu_1+\nu_2^1$ vibronic modes of CCH are strongly coupled
to the two lower A' components of the first $^2\Pi$ electronic excited state. The two A" upper
components of this state are strongly coupled to the ground vibrational levels $5\nu_2^1+\nu_3$
and $9\nu_2^1+\nu_3$. Hence, CCH has
a complex infrared spectrum below 12000 cm$^{-1}$
\citep[see, e.g.,][]{Carrick1983,Curl1985,Sharp2011} 
which is difficult to model as 17 $\Pi$ vibronic states are coupled at some level 
producing strong transitions to overtones and combination bands through infrared pumping
\citep{Tarroni2003,Tarroni2004}. \citet{Daniel2012}
have considered the effect of including or not the different vibrational modes of HNC in the radiative
transfer modelling. The bending mode, $\nu_2$, at 21 $\mu$m could be mainly populated from photons arising 
in the dust formation
zone while the stretching modes and their combination bands with the bending mode will be populated
through dust and stellar photons between 2 and 5 $\mu$m.

In order to quantify the role of stellar light variations we have developed a $RT$ model 
for HNC \citep{Daniel2012} with the abundance profile for this molecule in IRC+19216 derived 
recently from ALMA data by \citet{Cernicharo2013}. The results are shown on Figure 5. 
The left pannel show the relative influence of each 
vibrationnal mode on the emerging line intensities and profiles
for stellar phase $\phi$=0.3. The $\nu_2$ bending 
mode at 21\,$\mu$m clearly dominates in all observed rotational lines. For a molecule
having a large abundance in the innermost region of the CSE, which is not the
case for HNC, pumping through the stretching modes will be also very important (HCN for
example).
The right pannel show the intensities and profiles for different IR flux levels (i.e. stellar phases).
The predicted low-$J$ lines show
little change with the stellar phase, whereas high-$J$ lines show a strong intensity modulation. 
In spite of the simplicity of our model, which does not include time lags across the CSE (which
could be important for large CSEs and for short period stars), 
the predicted line profiles and intensities reproduce well the observations shown in Figures 2\&3. 
We also see that all $J/N$ lines behave differently: the $J$=1-0 line on Fig. 5 (right) is 
brighter at minimum IR intensity, whereas the lines with high $Js$ are dimmer. Indeed, IR pumping tends 
to populate the high energy $J$ levels at the expense of the lowest $J$=0, 1 and 2 levels. This is 
actually observed for both CCH $N$=1-0 and HNC $J=1-0$ 
on Figures 2\&3, which show that the light blue profile, corresponding to the minimum of IR flux 
(July 2012, see Figure 4), is brighter than all other stellar phases.

The question of different time lags between radiatively connected points of the 
envelope can partly be overcome when the velocity gardient is large with respect to 
turbulent velocity, which is the case for radii larger than
20R$_*$. A Large Velocity Gradient (LVG) decouples radiatively all, but the closest 
points in the envelope, drastically reducing the complexity of the $RT$ models.
As molecular radiation plays in this case only a local role we can solve those 
equations independently at every radius $r$. The stellar radiation field 
will be seen at distance $r$ from the star with a time lag $r/c$. The effect on the 
emerging line profiles
will depend strongly on the molecular line opacities. Optically thick lines arising from the whole envelope
as the low-$J$ lines of CO, HCN, SiC$_2$, will be less sensitive to IR pumping and show 
little fluctuation. However, the emerging high-$J$ lines, which are very sensitive to 
IR pumping and arise from smaller size regions, will also show a modulation with the stellar 
phase.

Finally, the blue and red horns of the cusped lines, that respectively arise from the front 
and rear caps of the expanding envelope, should exhibit a differential phase delay in their 
intensity variations, upon reaching the observer, that depends on the size of the envelope. 
This delay may be used to constrain the distance of CW~Leo, which is still poorly known 
\citep{Menten2012}. In the cases of CCH and HNC, which are mainly 
constrained within a couple of thin spherical shells of radius $\simeq$14$''$-20$''$ 
(corresponding to 3-5$\times$10$^{16}$ cm if the actual distance is 130 pc), a delay of 
20-30 days could be expected. Unfortunately, although several attempts
were done with the IRAM 30-m telescope, at one week intervals, to observe the lowest variable 
line of CCH, $N=4-3$, poor weather prevented us to obtain reliable data. 

In conclusion, the derivation of the physical parameters of variable star envelopes, from 
their sub-mm and far-IR molecular emissions, is not as straightforward as previously assumed. 
The intensities of sub-mm molecular lines of AGB star envelopes cannot be considered as reliable 
flux calibrators. Even the far-IR continuum emission from those envelopes may change significantly 
with the stellar phase \citep{Groenewegen2012}.
Nevertheless, the low-$J$ lines of CO, which have been used to
derive the mass loss rate of AGB stars, are not significantlly affected by their periodical
luminosity variations. 
The interpretation of sub-mm and far-IR line observations of CSEs
%, in particular of molecular line data, 
requires to elaborate $RT$ models that include the IR flux variations and IR pumping
throughout the envelope. Resulting changes in the dust and gas temperatures
in the innermost layers of the CSE have also to be implemented.
Each molecule in this game has its own peculiarities (abundance profile, frequency and intensities
of the vibrational modes). The spatial extent will also affect the balance between collisional
and IR pumping.
Moreover, these effects
could be different for each isotopologue of a given molecule
due to the different opacities of their rotational and ro-vibrational lines.

\acknowledgements
JC, MA, GQL, and LVP thank spanish MICINN for funding 
under grants AYA2009-07304, AYA2012-32032, CSD2009-00038, and 
ERC under ERC-2013-SyG, G.A. 610256 NANOCOSMOS.
LD acknowledges financial support from the Fund for Scientific Research - Flanders (FWO).

\end{document}